# Distal edge determination precision for a multi-slat prompt-gamma camera: a comprehensive simulation and optimization of the detection system

A. Morozov[1,*], H. Simões[1] and P. Crespo[1,2]

[1]LIP-Coimbra, Departamento de Física, Universidade de Coimbra, 3004-516 Coimbra, Portugal

[2]Departamento de Física, Universidade de Coimbra, 3004-516 Coimbra, Portugal

*Email: andrei@coimbra.lip.pt

**Abstract**

Multi-slat prompt-gamma camera is a promising tool for range monitoring during proton therapy. We report the results of a comprehensive simulation study analyzing the precision which is possible to reach with this camera in determination of the position of the distal edge of the Bragg peak. For the first time we include simulation of optical photons. The proton beam (single pencil beam, 130 MeV, 10 ns bunch period, total of $1 \cdot 10^8$ protons) is interacting with a polymethyl methacrylate (PMMA) phantom, which is a cylinder of 200 mm in diameter and length. The prompt gamma rays generated in the phantom are collimated with a multi-slat collimator and detected using a combination of yttrium aluminum perovskite (YAP) scintillators, installed in the collimator apertures, and light sensors. Two scintillator packing schemes, with one and with two scintillator plates per aperture, are considered. The collimator configuration (the septal thickness, aperture and height), resulting in the best precision, is determined using two methods of detector optimization. Precision of 2.1 mm (full width at half maximum) in the edge position determination is demonstrated.

# 1    Introduction

The main advantage of proton therapy (PT), in comparison to the conventional radiotherapy with megavoltage X-rays, is that a proton beam can deposit energy in a well-localized region inside the patient's body (the Bragg peak region) with a sharp distal dose fall-off [1]. Due to this property, PT can deliver a highly conformal radiation dose to a tumor while simultaneously minimizing the integral dose and sparing surrounding healthy tissues.

However, the highly localized dose delivery comes with a price: even a small (on the order of a few millimeters) error in the proton beam range can lead to a significant under-irradiation of the tumor or over-irradiation of the healthy tissue of closely situated vital organs [2]. The proton range strongly depends on tissue composition, density, and heterogeneity. The main causes of range uncertainties are related to the conversion of X-ray computed tomography data to proton interaction data as well as morphological, anatomical and physiological changes in the patient's body occurring during the course of the therapy [3].

Several approaches have been suggested for in vivo monitoring of the PT dose delivery and proton beam range verification. One of them, already applied in clinical practice, is based on positron emission tomography (PET) which involves detection of 511 keV gamma rays resulting from positron-emission decay of proton induced radioactive nuclides [4].

Another group of techniques is based on detection of prompt gamma rays (PG) originating from proton-nuclear interactions within the body [5]. The PG-based techniques do not rely on detection of delayed emission, as the PET-based ones do, and, therefore, are unaffected by biological processes such as the activity washout [6], which is a significant advantage targeting accurate range monitoring. This group of techniques can be sub-divided into prompt-gamma imaging (PGI), prompt-gamma timing [7] and prompt-gamma spectroscopy [8].

There are two main approaches in PGI: the first one is to use a collimator-less system based on the Compton camera principle (see e.g. [9, 10]), while the second one is to use a combination of a collimator and a gamma ray detector (or an array of detectors). The collimation, in turn, can be organized either using the "camera obscura" principle (e.g., pinhole camera [11] and knife-edge shaped slit [12, 13, 14]) or applying orthogonal imaging, with, for example, a single slit [15, 16], a multi-slit (an alternative name is multi-slat) [17, 18]) or a multi-hole [19] collimator.

Several PGI techniques are currently under active development, showing a potential to reliably identify range shifts of about 2 mm [16, 20]. First clinical application of a PGI-based technique with a knife-edge shaped slit camera has already been reported [13]. Further details on the current status and perspectives of the range monitoring in proton therapy can be found in a recent review article [21].

A PGI imaging system with a multi-slat collimator oriented orthogonality to the beam direction has been investigated in our previous studies [18, 22]. Simulated PG profiles obtained with six different multi-slat collimators were compared to determine which collimator configuration works better considering the precision of the distal edge position determination. The detector system was not simulated on purpose, in order to give emphasis on the effect of the collimator itself. Pencil proton beam irradiation of the head (130 MeV proton energy) and pelvis (200 MeV) of an anthropomorphic phantom was simulated with and without physiologic/morphologic or setup changes of clinical dosimetric relevance. The particles escaping the phantom were transported through each of these multi-slat configurations and the gamma count profiles were recorded at the collimator exit. Median filtering was applied to the registered PG-profiles to mitigate the effects of septa shadowing and statistical fluctuations. Time-of-flight discrimination was used to enhance the signal-to-background ratio. It was reported that visual detection of the artificially introduced changes is possible. The results have also demonstrated that 2 mm range shifts could be detected for the head irradiation scenario using a simple linear regression fit to the fall-off of the PG profiles.

In this paper we present the results of a follow-up study dedicated to the multi-slat prompt-gamma camera. According to [20], this type of imaging systems can have an advantage in the precision of the distal edge determination compared to the knife-edge based cameras. The main goal here is to conduct a comprehensive simulation/optimization of the entire detection system as the first step towards construction of an experimental prototype. The study includes simulation of generation, transport and detection of optical photons, which, to the best of our knowledge, has not been reported before. We have replaced the ideal particle detectors used in the previous simulations with a model of scintillators interfaced to light sensors. For the scintillation material we have selected cerium doped yttrium aluminum perovskite (YAP) due to an attractive combination of properties: good photon yield (17 photons per keV [23]), fast scintillation (0.38 ns rise and 27/140 ns decay times [23]), absence of intrinsic radioactivity and non-hygroscopicity of the crystal.

In this study we evaluate the precision in the determination of the distal edge position for 130 MeV proton beam. Only a simple phantom, a polymethyl methacrylate (PMMA) cylinder, was considered here, thus leaving investigation of the edge determination precision for anthropomorphic phantoms to a follow-up study. Two approaches to determine the distal edge position were developed and compared. The first one is based on analysis of the energy deposition inside the scintillators. This method can provide the ultimate precision but it is applicable only for analysis of simulation results, when the deposition energy, time and position are known for each deposition node. The second method involves generation and analysis of signal waveforms of the optical sensors, and thus is applicable for analysis of experimental data.

We also present the results of an optimization of the collimator design, targeting the highest possible precision in the edge determination. The optimization was performed over three parameters: collimator's septal thickness, aperture and height. Additionally, two packing schemes of the scintillators were considered: with one and with two scintillator plates per collimator aperture.

## 2 Methods

### 2.1 Simulation tools

Monte Carlo simulations of particle transport were performed using the Geant4 toolkit [24, 25, 26] version 10.6.1, which was the newest version at the time of the study. The QGSP_BIC_HP reference physics list was activated, enabling the high-precision models for both electromagnetic processes (Option4) and neutron transport. The production cut was set to 0.5 mm for the phantom region and to 0.05 mm for the rest of the setup.

The ANTS2 toolkit [27] version 4.31 was used as the front-end for Geant4. ANTS2 provides a scripting/GUI interface for interactive definition and visualization of the detector geometry. It also offers a collection of tools for storage and analysis of the simulation results, and gives access to the semi-automatic optimization procedure of the detector design applied in this study (see section 2.6).

Wavelength-resolved simulations of generation, transport and detection of optical photons were performed in ANTS2. The toolkit takes into account time- and wavelength-related properties of the scintillation and light detection, as well as photon interaction properties for the material bulk and the material interfaces.

### 2.2 Proton beam

The proton beam was simulated as a single pencil beam of zero width. The proton energy was set to 130 MeV, corresponding to a typical head irradiation scenario [22]. One PT *frame* was simulated with $1 \cdot 10^8$ protons (322580 bunches of 310 protons each, bunch period of 10 ns), which is representative of the most distal pencil-beam in a typical cyclotron-based treatment plan [28]. The overall duration of one frame is thus about 3.23 ms. All protons in a bunch had the same generation time (bunch duration is zero).

In the following text, the *global time* refers to the time elapsed from the start of the beam. The *time from bunch* refers to the time elapsed from the previous bunch (the global time modulo bunch period, therefore assuming values in the range from 0 to 10 ns).

## 2.3 Simulated configurations

All particle transport simulations were conducted in this study with a setup consisting of two parts: the phantom and the camera. The phantom was a PMMA cylinder of 200 mm length and 200 mm in diameter, representing the patient's head. The camera was composed of a multi-slat tungsten collimator and YAP scintillators, inserted between the collimator plates (see figure 1). The distance between the surface of the PMMA cylinder and the top plane of the collimator was set to 150 mm.

In all simulations the proton beam was irradiating the phantom at the center of one of the cylinder's faces along the axis of the cylinder. We assume X axis is along the proton beam direction and Z axis is orthogonal both to the beam direction and the top surface of the collimator.

The collimator sizes along X and Y directions were fixed to 200 mm and 400 mm, respectively. The collimator size in Z direction, as well as its septal thickness and aperture size, were optimized in this study. The values for these parameters were confined to the ranges of 100 – 300 mm, 1 – 6 mm and 1 – 6 mm, respectively.

The scintillator plates had a fixed Z size of 30 mm. Along Y direction, the scintillator plates had the size equal to that of the collimator (400 mm). Each scintillator plate was split in 15 equal segments, resulting in Y size of a single scintillator segment of 26 mm (see figure 1(b)).

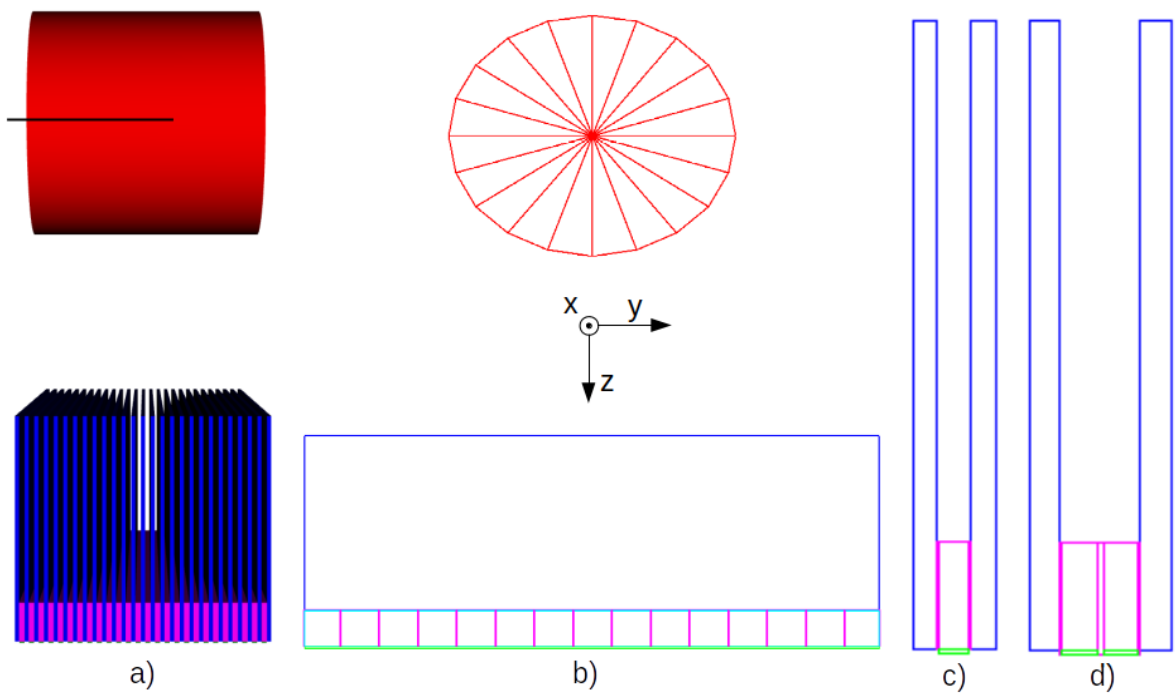

Figure 1. a) Perspective view of the setup along Y direction. The thick black line represents the proton beam. b) Parallel view (wire-frame) of the setup along X direction. c) and d) Parallel view along Y direction, showing one elementary cell of the collimator for configurations with one (c) and with two (d) scintillators per aperture. The phantom is red, the collimator blades are blue, the encapsulated scintillators are magenta and the optical sensors are green.

Two schemes of packing of the scintillators were considered. In the first one a single scintillator plate, and in the second scheme two scintillator plates, placed side by side, were occupying each aperture (see figure 1(c) and 1(d)). Each scintillator was surrounded by a 0.275 mm layer of polytetrafluoroethylene (PTFE) encapsulation on all sides except one, where it was interfaced to a 1 mm slab of epoxy representing the window of a light sensor.

For development and validation of the model used to generate signal waveforms of the light sensors in response to an energy deposition event inside a scintillator, the following configuration was developed in ANTS2. A single YAP scintillator with the dimensions of 26 x 3 x 30 mm$^3$ was interfaced through a 0.2 mm thick epoxy layer to an ANTS2 light sensor (see figure 2). On all other sides the scintillator was surrounded by a 0.25 mm thick layer of air. The material of the "world" object was either air or PTFE.

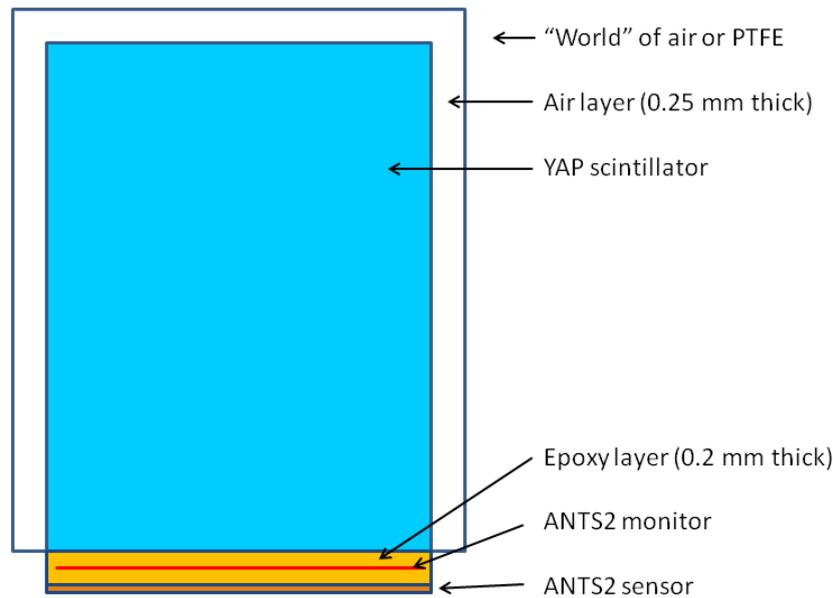

Figure 2. Model of the scintillator segment interfaced to a light sensor. The scintillator slab has the dimensions of 26 mm width, 30 mm height and 3 mm depth (direction along the viewing axis). It is surrounded on all faces, except the lower one, with a layer of air (0.25 mm thick). The lower face is in contact with an epoxy layer, containing an ANTS2 monitor. The epoxy layer is interfaced to an ANTS2 sensor. The material of the world object was either air or PTFE. Relative sizes are not to scale.

The wavelength-resolved optical properties were taken from the following sources: the refractive index and attenuation length for YAP from [29], YAP emission spectrum from [30], the refractive index for epoxy from [31] (resin "8") and the refractive index for PTFE from [32] (grade AF1601). Attenuation in the epoxy layer was ignored. Note that there is a significant self-absorption for the short-wavelength YAP emission.

The epoxy slab contained an ANTS2 *monitor* object. The monitor was used to record time of arrival of optical photons at the ANTS2 light sensor. The sensor was used to record the overall signal, taking into account the wavelength-resolved photon efficiency given by the Hamamatsu company for the short-wavelength avalanche photodiode model S8664-55 [33]. Note that this sensor model has an epoxy entrance window.

In order to simulate YAP encapsulation by PTFE, a special rule for the air-to-PTFE interface was defined in ANTS2 according to the *FSNP* model described in [34]. In short, when a photon arrives at the interface, first a Fresnel equation is used to determine the specular reflection probability. If reflection is not triggered, it is assumed that the photon enters the PTFE bulk. A check against albedo is made then, and if it fails, the photon is absorbed. Otherwise, the photon is back-scattered from the interface with the direction sampled according to the Lambert law. We assumed the PTFE albedo of 0.93 [35]. Note that before exiting YAP to the air layer, a Fresnel reflection is checked, so full internal reflection on this interface is taken into account. This is an important process due to high (close to 2) refractive index of YAP. A cross-test validation of the ANTS2 model was performed using Geant4 and considering a YAP slab with an optical surface implementing the DAVIS model with dielectric_LUTDAVIS type and PolishedTeflon_LUT finish. Both models gave similar results on the photon collection efficiency as a function of the source position.

## 2.4 Simulation approach

Simulations were conducted in three independent stages. In the first stage interaction of the protons of the beam with the phantom was simulated. All particles leaving the phantom in the direction of the collimator during a time window equal to the beam duration (3.23 ms) were recorded in a binary file.

In the second stage a simulation was performed to obtain energy deposition in the scintillators. The records of the "primary" particles were loaded from a binary file generated during the first stage. Deposition data (deposited energy, position, time and particle type) were saved in an output binary file.

In the third stage the deposition information obtained in the second stage was used to generate signal waveforms (1 ns time bin, 3.23 ms duration) of the light sensors. Tracing of optical photons for each deposition event was found to be inefficient (very time consuming). As an alternative, a mathematical model was developed which was used to generate contribution to the signal waveform of a light sensor based on the parameters of the deposition event (see section 3.5).

## 2.5 Edge determination procedure

Two procedures for determination of the edge position were developed in this study. The first one, described in this section, uses information on the energy deposition available in the simulations (position, time and energy for each deposition *node*). This approach cannot be applied for analysis of experimental data, but the results obtained with this method using simulation data can be considered as the best-case scenario for less direct methods. Another procedure, based on analysis of signal waveforms generated by the light sensors is described in section 3.6.

Geant4 simulations of a PT frame (stage two, see section 2.4) were used to build lists of energy deposition nodes (the deposited energy and the global time of the deposition) for each scintillation plate. The global time was converted to the time from bunch (global time modulo bunch period), and a filtering based on this value was performed discriminating non-prompt events. For the considered configurations of the camera the typical acceptance window had a range from 1 to 3 ns (see section 3.2). The deposition data were used to build profiles of the deposited energy as a function of the scintillator plate index (see an example in figure 3).

The profiles were used to extract the distal edge position. First a median filter (kernel of 3 and 6 bins for the configurations with one and two scintillator plates per collimator aperture, respectively) is applied to the profile to suppress the outliers. Then the bin with the maximum deposition energy value is found. Next, moving from this bin towards larger scintillator indexes, the bin is found where the deposited energy becomes less than a threshold value, equal to a certain fraction of the peak value. The position corresponding to the distal edge is calculated using linear interpolation of the values and positions of two neighboring bins with the deposition energy above and below this threshold. In this study we have found that the best precision is obtained with the threshold fraction of 0.5 of the peak.

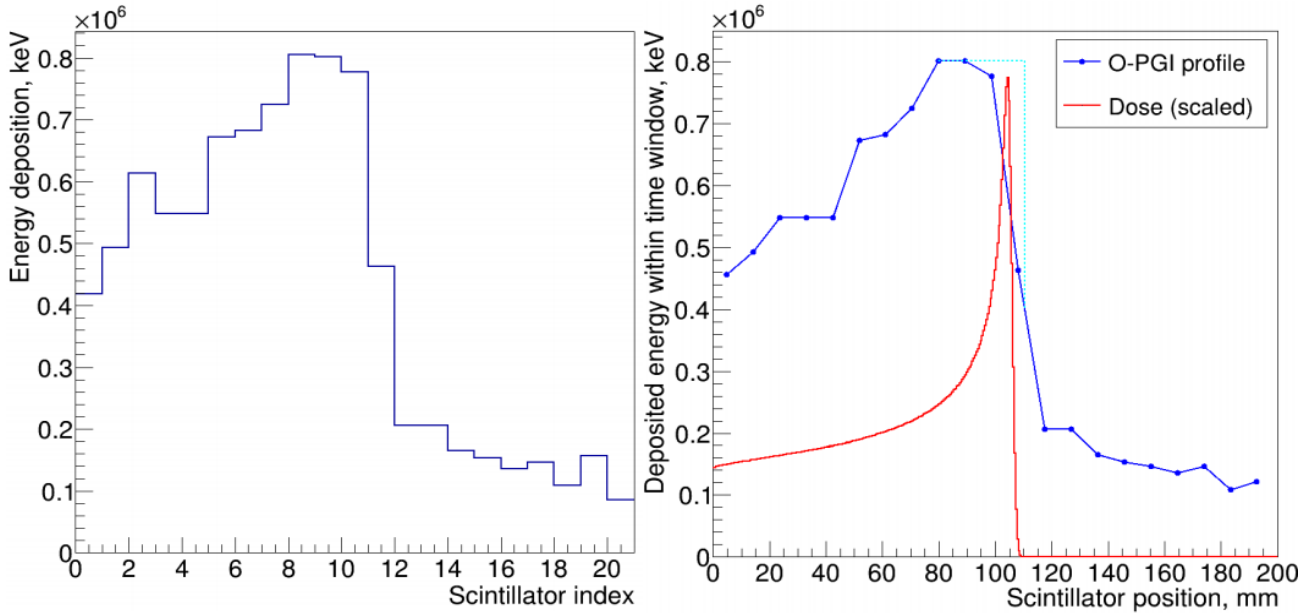

Figure 3. Left: Example of a profile of the deposited energy as a function of the scintillator plate index. Right: The same profile after applying the median filer and converting the scintillator index to the corresponding X position (center of the slab). The vertical dotted line indicates the extracted edge position. The red curve shows a scaled profile of the dose inside the phantom along the beam direction.

Note that the computed value, which is declared to be the edge position has an offset from the true position. As this study is focused on the method precision (reproducibility of the computed position), calculation of this systematic offset was not performed.

## 2.6 Collimator design optimization

Optimization of the camera design is not a trivial task due to a combination of several factors. Firstly, the edge computation results are quite noisy: fluctuations in the determined edge position from simulations different only by the seed of the random number generator can be quite strong. Secondly, the edge determination procedure should perform well for the entire range of relative positions of the Bragg peak in respect to the position of the nearest collimator blade. These two factors require to perform a large number of tests for every configuration to reliable determine how good it is in terms of the edge determination precision. Thirdly, the simulation time per configuration is quite long. Even splitting the simulations in stages (section 2.4) in order to re-use the results of the most time consuming first stage, it takes about 20 minutes with a single thread on a desktop PC with i7-7700K (4.2 GHz) processor to perform the second stage simulation of one PT frame.

Assuming normal distribution of the extracted edge positions, simulation of about 50 frames is required to reach a standard error of about 10% in the standard deviation of the determined edge position:

$$SE(\sigma) = \frac{\sigma}{\sqrt{2N-2}}$$

Simulation of 50 frames using 7 threads takes about 3 hours. This means that a brute force optimization approach with 3 parameters (the collimator's septal thickness, aperture and blade length) becomes a very time consuming procedure even considering a quite sparse grid of nodes in the parameter space.

In this study we have decided to perform optimization of the design using two different approaches. The first one is based on a semi-automatic optimization using the infrastructure already available in ANTS2 [36]. The main idea is to implement a standard optimization algorithm (e.g. simplex) and provide a custom functor which automatically evaluates a custom cost function for a given configuration of the camera based on the parameter values provided by the optimizer at each iteration.

The functor was developed in ANTS2 using the scripting system of the toolkit. At each iteration it receives the values of the free parameters from the optimizer, rebuilds the camera geometry using these values, performs a simulation of a given number of frames and determines the edge position for each of them. The cost function, returned to the optimizer, was defined as the standard deviation of the corrected edge positions for these frames. The corrected edge position is defined as the computed edge position for a given frame minus the phantom shift used in that frame simulation. Obviously, the smaller is the sigma, the better is the extraction method precision. The optimization runs were performed with the total number of 120 frames at each iteration (see next section).

As an alternative, a brute-force optimization was conducted delegating the second stage simulations to the Navigator cluster [37] of the Laboratory for Advanced Computing of the University of Coimbra. The edge determination accuracy was evaluated only for the camera design with one scintillator per collimator aperture. A total number of 1331 configurations were investigated: the collimator septal thickness from 1.0 to 5.0 mm, the blade length from 100 to 300 mm and the aperture from 1.0 to 5.0 mm (11 nodes in each dimension with regular step). Simulations of 60 PT frames were performed for each configuration.

# 3 Results

## 3.1 Simulation: phantom

To generate sufficient data for evaluation of the precision which can be obtained with the multi-slat prompt-gamma camera, 120 first stage simulations were conducted in total with 12 different X positions of the phantom's entrance face, regularly spaced in the range from 0 to 8.25 mm with a 0.75 mm step. For each position 10 simulations were performed with different seeds of the random number generator. As the collimator's position was fixed, a given phantom shift should result in an equal shift in the determined position of the Bragg peak edge. One simulation, performed in 7 threads, took about 3 hours on a desktop PC with an i7-7700K processor.

Statistics on the secondary particles leaving the phantom in the direction of the collimator for one PT frame ($1 \cdot 10^8$ primary protons) is given in table 1, and the particle energy spectra are shown in figure 4.

| **Particle** | **Number (per primary particle)** | **Mean energy [MeV]** |
|---|---|---|
| Gamma ray | $2.46 \cdot 10^{-2}$ | 2.7 |
| Neutron | $1.58 \cdot 10^{-2}$ | 13.7 |
| Electron | $4.3 \cdot 10^{-4}$ | 2.3 |
| Proton | $5 \cdot 10^{-5}$ | 20 |
| All other | $4 \cdot 10^{-5}$ | - |

Table 1. Number (per primary particle) and the mean energy of the secondary particles leaving the phantom in the direction of the collimator for one PT frame with the proton energy of 130 MeV.

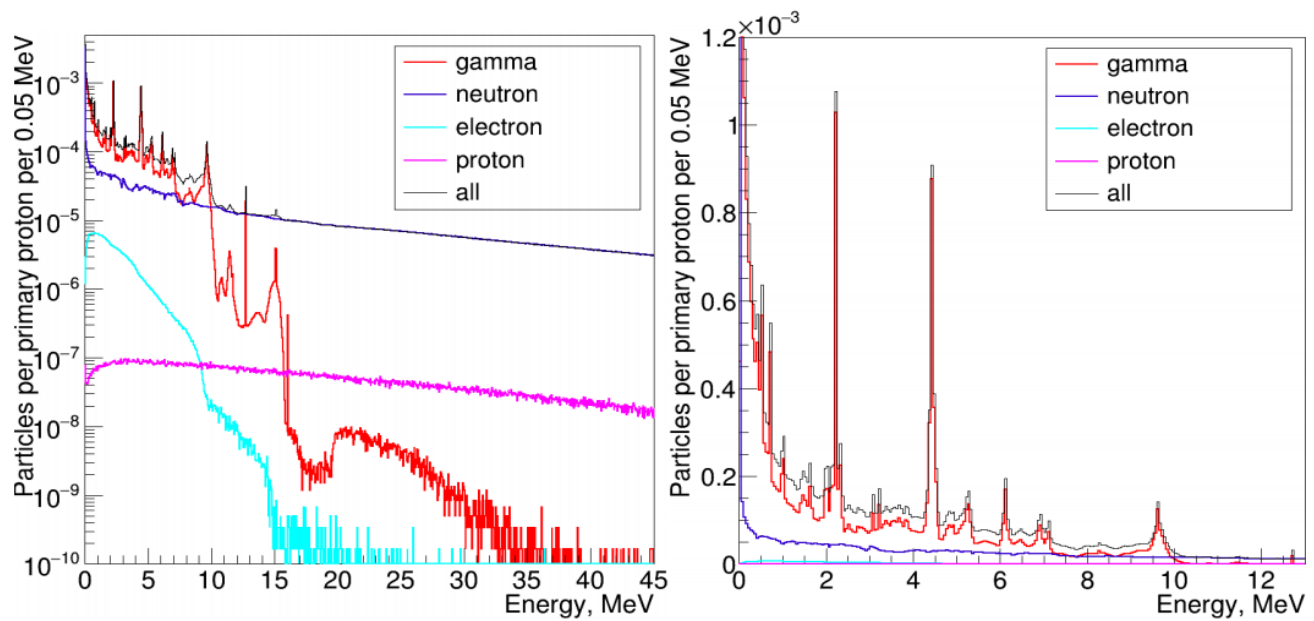

Figure 4. Energy spectra of the secondary particles leaving the phantom in the direction of the collimator. The curve titled "all" represents the data accumulated over all particles leaving the phantom.

The results show that the number of generated neutrons is large (about one neutron per two gamma rays), and the mean energy of neutrons exceeds 10 MeV. Due to frequent scattering in the phantom and the collimator, neutrons (and, consequently, the secondary particles generated in neutron interactions) do not carry spatial information correlating with the dose distribution in the phantom and thus result in a background in the output, having a negative effect on the precision of the edge determination (see next section).

The simulation model was benchmarked against previously reported results [20, 38] for a cuboid PMMA phantom (600 x 100 x 100 mm$^3$) and proton energy of 134 MeV. The obtained yields of secondary particles per primary proton are shown in Table 2 together with the values from [20] and [38] (older versions of Geant4 as well as the results obtained with Fluka code [39]). Note that the results are quite similar, and the gamma-to-neutron ratio is close to unity for all cases.

| Secondary particle | Yield of the secondary particles per primary proton | | | |
|---|---|---|---|---|
| | This work | [20], Geant4 | [38], Geant4 | [38], Fluka |
| Gamma (>1MeV) | 0.0706 | 0.0869 | 0.0945 | 0.0493 |
| neutron | 0.0897 | 0.0823 | 0.0871 | 0.0555 |
| proton | 0.00108 | 0.00120 | 0.00129 | 0.00104 |

*Table 2. Yields of the secondary particles per primary proton obtained in this work and in [20] and [38] for proton energy of 134 MeV and a cuboid phantom of PMMA (600 x 100 x 100 mm$^3$).*

## 3.2 Simulation: multi-slat prompt-gamma camera

An example of the results of the second stage simulations is given in figure 5. The amount of energy (color-coded) deposited in each scintillator plate during a PT frame is shown as a function of time from bunch (global time modulo bunch period) and scintillator index. This is the most meaningful time scale as the prompt gamma rays originating from the phantom interact with the scintillators after a fixed delay on the order of 1 ns from the bunch time. Due to collimation (each scintillator can directly "see" only a section of the phantom), this delay systematically shifts to larger values with increase of the scintillator index due to longer flight time of the primary protons inside the phantom before they enter the field of view of the further-positioned scintillators.

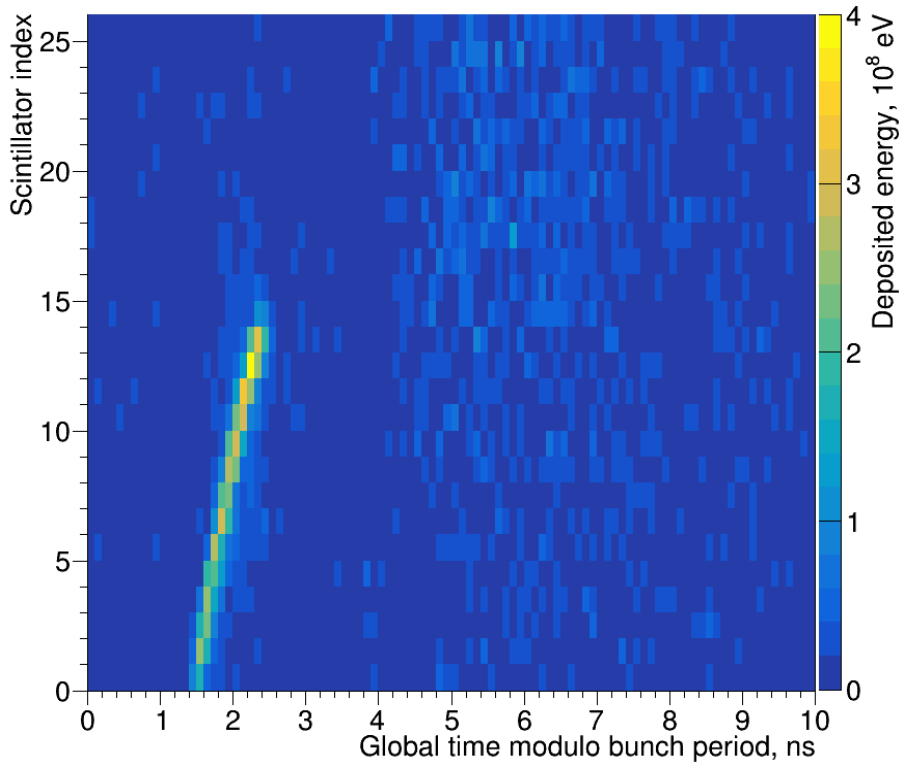

Figure 5. Map of the deposited energy (color-coded) during a PT frame vs time from bunch and scintillator index. The deposited energy is given in the units of eV per scintillator plate per 1 ns.

The strong deposition component appearing in the time range from 1 to 3 ns is due to the prompt gamma rays. Neutrons leaving the phantom can have significantly longer flight time inside the phantom, in the phantom-to-collimator gap and inside the collimator. Therefore, deposition from the products of interactions of neutrons with the phantom/collimator/scintillators can appear with a significant delay from the bunch time, resulting in a broad continuum (longer than the bunch period).

Figure 6 shows two projections of the map of figure 5: along the time axis and along the scintillator index. The deposition vs time profile clearly shows both the prompt and the delayed components. The spatial profile exhibits a drop after scintillator index 13, which is associated with the end of the prompt gamma ray continuum. However, the prompt gamma profile is superimposed over a very strong and noisy background.

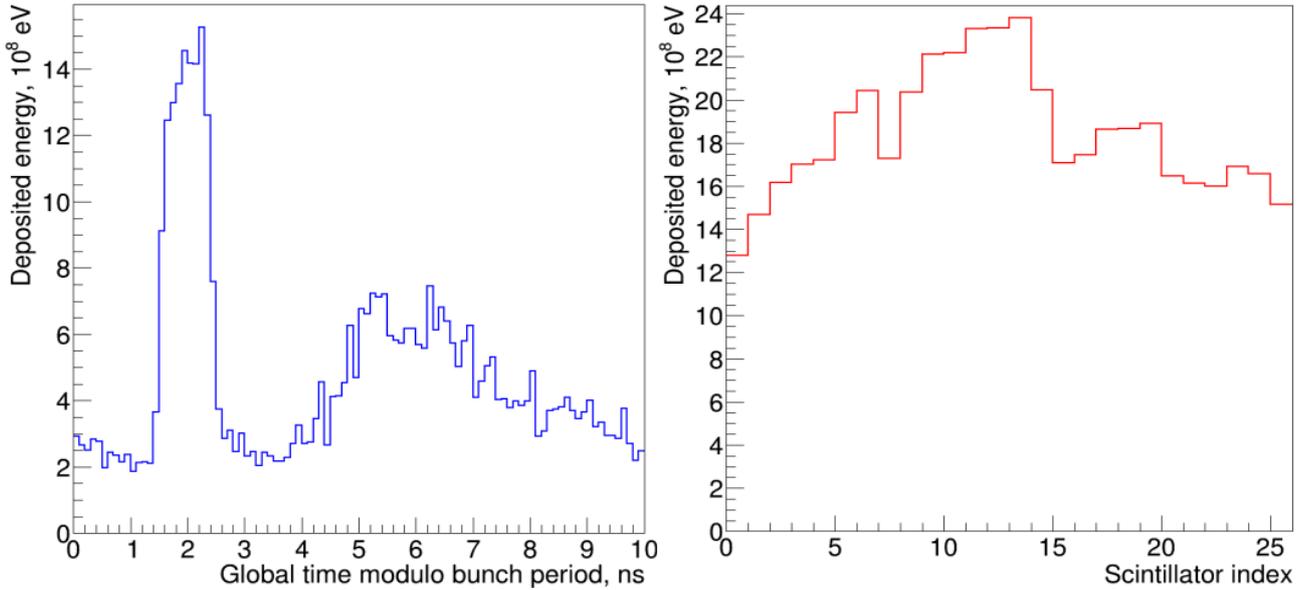

Figure 6. Deposited energy during one PT frame vs time from bunch (left) and vs scintillator index (right).

The background level in the spatial profiles can be reduced using two methods [22]. The first one is to introduce time-based discrimination. Disregarding the deposition events appearing outside of a time window from 1 to 3 ns can significantly improve the contrast between the prompt and slow components (see figure 7). Introduction of this filtering is feasible considering practical implementations since YAP scintillator has very good timing properties: sharp 0.38 ns rise time and decay times of 40 ns (89%) and 140 ns (11%) [23].

The second method is to introduce event discrimination based on amount of deposited energy. Figure 8 show time profiles of events with total deposited energy inside a scintillator above and below a given threshold. For low thresholds (about 250 keV) a large fraction of the slow component is suppressed, while the reduction of the prompt component is negligible. With increase of the threshold, the reduction factor of the slow component start to increase slower, while reduction of the prompt component becomes significant. We have selected a threshold value of 750 keV, resulting in an acceptable reduction of the prompt component of about 20% but providing nearly 50% reduction in the background level (see figure 8). The reduction in the background level in the profile of the deposited energy vs scintillator index obtained by introducing this energy filter is demonstrated in figure 7 (thick red curve).

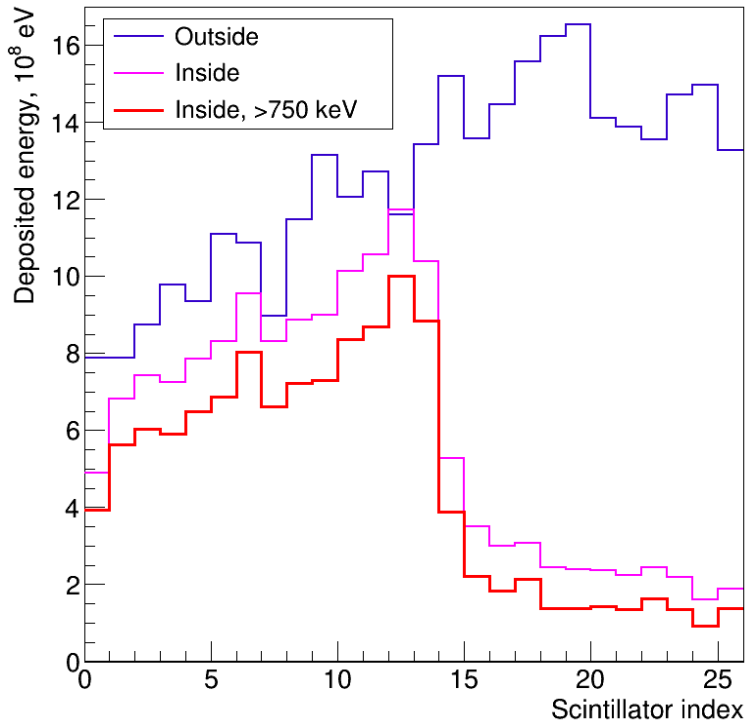

*Figure 7. Profiles of the deposited energy vs scintillator index considering only depositions with the time-from-bunch inside (magenta) and outside (blue) the range from 1 to 3 ns. The red thick line shows the profile obtained with both time (considering only depositions within 1 - 3 ns) and energy (considering only depositions above 750 keV) filtering.*

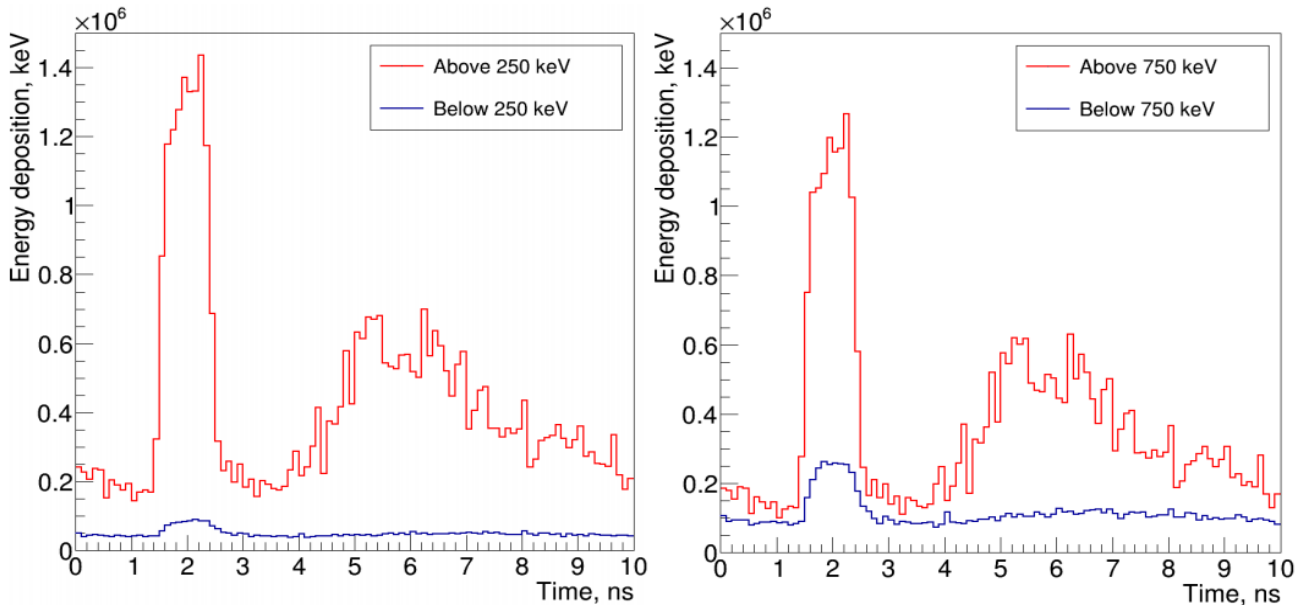

Figure 8. Profiles of the deposition energy vs time from bunch, above and below the energy threshold for two values of the threshold (250 and 750 keV).

Note that the energy-based discrimination was performed independently for results of each individual beam bunch. Due to tracing of secondary electrons and photons, the output of Geant4 simulations often gives many individual deposition nodes with position and time very close to each other. Therefore, before performing discrimination, the deposition nodes situated inside one scintillator were combined if the difference in time between them was less than 0.1 ns. Note that a more realistic approach to this "clustering", taking into account overlaps in energy deposition from different bunches, is performed with the waveform-based edge position determination method, described in section 3.6.

## 3.3 Optimization: semi-automatic approach

Optimization of the camera design was performed for both schemes of scintillator packing (with one and with two scintillator plates per aperture of the collimator). The optimization procedure, described in section 2.6, was searching for the combination of the collimator's septal thickness, blade height (Z direction) and aperture width (X direction) that results in the smallest standard deviation (σ) in the extracted edge positions, corrected for the phantom shift. Several optimization runs were performed for both configurations using different initial conditions.

The best σ of 0.85 ± 0.05 mm was found for the configuration with one scintillator per aperture. An example of the iteration history (σ vs iteration number) is given in figure 9. That optimization run has started with the initial conditions of 1 mm septal thickness, 150 mm blade height and 1.5 mm aperture. The optimization results suggest the presence of a broad plateau with σ ≤ 1.0 mm for the parameter values of about 4.0 mm septal thickness, 190 mm collimator height and 4.0 mm aperture. Note that the peak at iteration #3 of figure 9 appears due to the workflow of the simplex algorithm: the first step (iteration #0) is used to compute σ at the configured initial values of the free parameters. Since there are three free parameters, the next three iterations are performed by adding the initial step value to each of the parameters' initial values. Thus the actual optimization starts only from iteration #4.

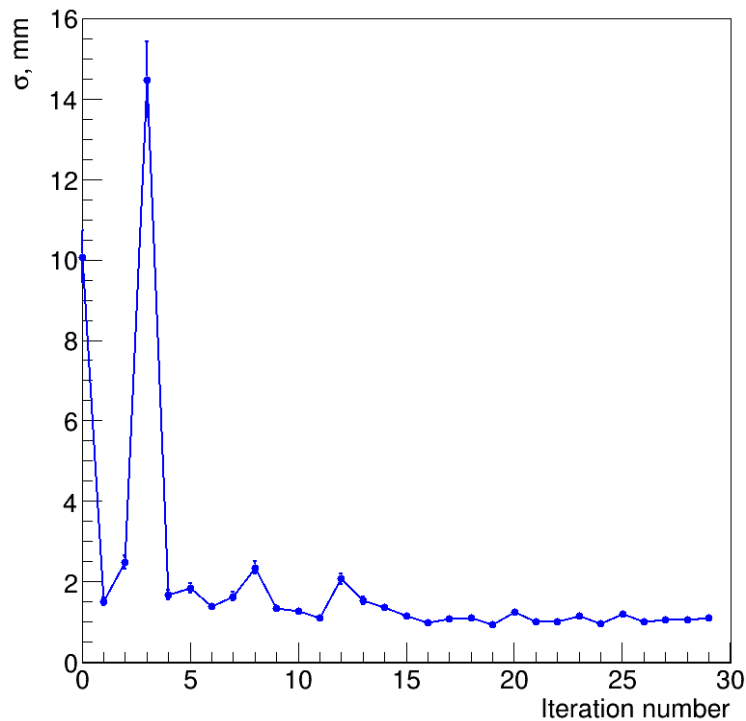

Figure 9. Example of the optimization history for the configuration with one scintillator per aperture. 120 frames were simulated per iteration. The minimum value of sigma is 0.85 ± 0.05 mm.

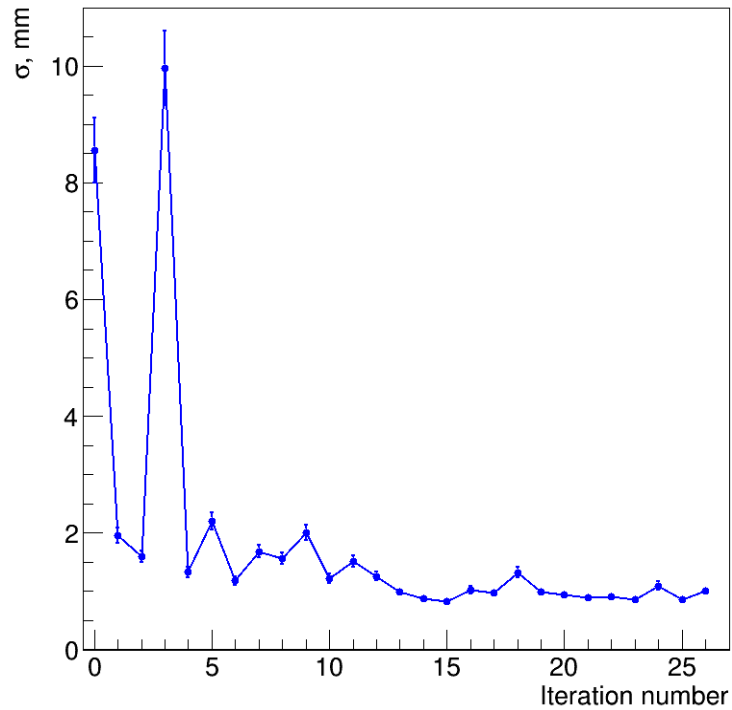

Figure 10. Example of the optimization history for the configuration with two scintillators per aperture. 120 frames were simulated per iteration. The minimum value of sigma is 0.96 ± 0.06 mm.

For the configuration with two scintillators per aperture the best σ of 0.96 ± 0.06 mm was found. An example of the iteration history is shown in figure 10 for optimization run with the initial conditions of 1.5 mm septal thickness, 120 mm blade and 2 mm aperture. A broad plateau with σ ≤ 1.20 mm was found for the parameter values of about 2.5 mm septal thickness, 250 mm collimator height and 4.0 mm aperture.

## 3.4 Optimization: brute-force approach

The brute-force optimization procedure, described in section 2.6, was performed only for the packing scheme with one scintillator plate per aperture. The computation time on the cluster for 1331 configurations totaled to 22000 hours. The obtained results confirmed the conclusion based on the results of the semi-automatic optimization: there is a broad plateau with the sigma value of about 1.0 mm for the configurations with the septal thickness of 3.0 – 5.0 mm, the blade length of 140 – 200 mm and the aperture of 2.5 – 5.0 mm.

Two examples of the color-coded maps of the standard deviation (σ) of the computed edge position, corrected for the phantom shift, versus septal thickness and aperture, are shown in figure 11 for the collimator blade lengths of 140 and 180 mm. The standard error is about 10% of the sigma value.

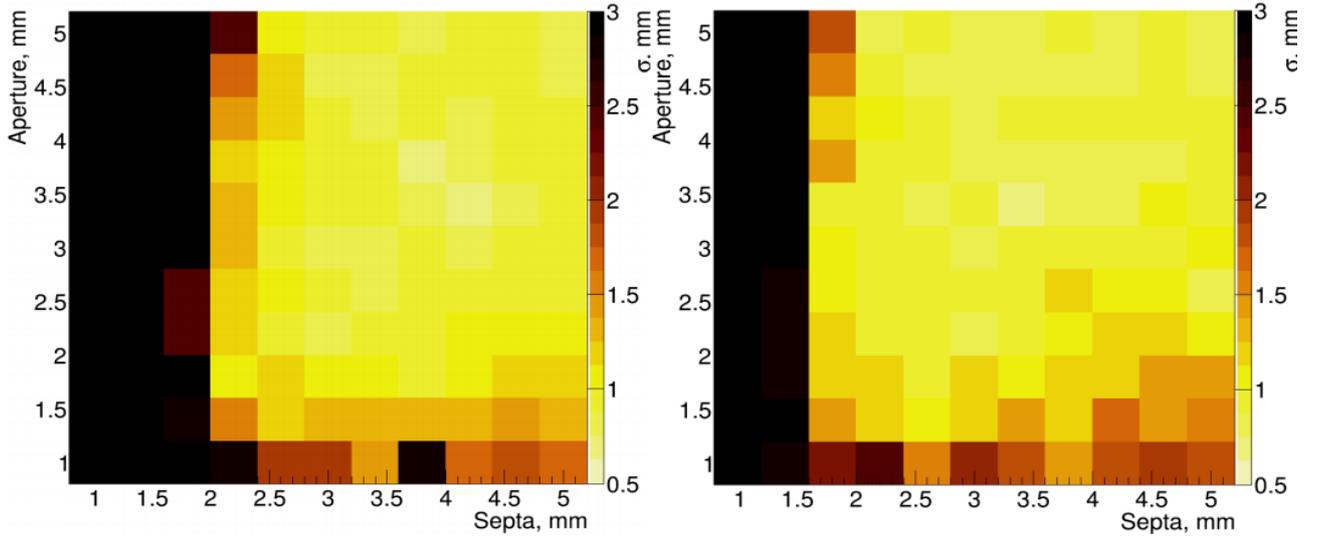

Figure 11. Color-coded maps of the standard deviation (σ) of the computed edge position, corrected for the phantom shift, versus septal thickness and aperture, for blade lengths of 140 mm (left) and 180 mm (right).

## 3.5 Optical simulations and sensor signal generation

This section describes the procedures developed for generation of the signal waveforms of the light sensors based on the energy deposition in the scintillators. We have decided to use 1 ns binning for the waveforms. A smaller time bin may seem to be better suited in order to reach higher accuracy in determination of the deposition time during processing of the waveforms, but considering strong increase in the statistical fluctuations due to a relatively small number of photons detected per bin (see section 3.6), 1 ns binning seems to be the best compromise.

The full-system simulations, in which optical photons are generated and traced for every energy deposition event were found to be inefficient due to a very large (~$1 \cdot 10^9$) number of photons generated per PT frame. Tracing this number of photons takes more than an order of magnitude longer time compared to the simulation of particle transport inside the camera.

A more efficient alternative was applied in this study. We have built, verified and applied a model which provides time-resolved contribution to the signal waveform based on the deposited energy, the global time and the position of the deposition node inside the scintillator.

Using the ANTS2 model of the scintillator slab with the attached light sensor (see section 2.3 and figure 2), the distribution of photon travel time from the emission point to the light sensor was obtained for several source positions inside the scintillator. The results (see figure 12) clearly show the presence of the direct and the 1st reflected components: a large fraction of the photons emitted in the direction away from the sensor travels until the opposite side of the scintillator and, after reflection and travel back, is detected by the sensor with a considerable (~0.3 ns) delay. Higher order reflection components are also present, but their combined effect is negligible in comparison with the total number of detected photons.

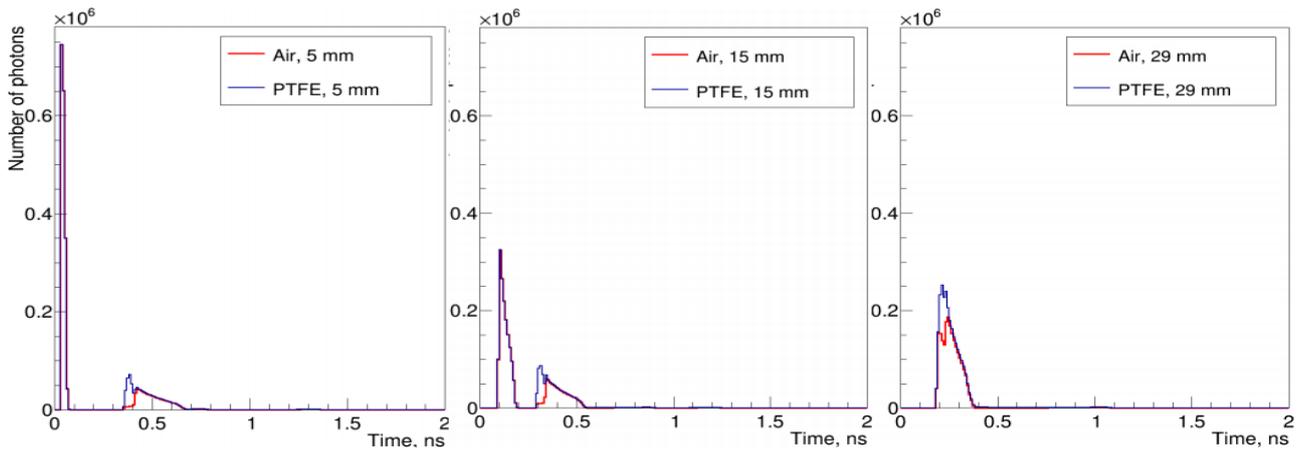

Figure 12. Distribution of photon time-of-flight from emission until arrival at the sensor. The time bin is 0.01 ns and the total number of emitted photons is $1 \cdot 10^7$. The three graphs show the distributions for the distances of 5, 15 and 29 mm (from left to right) between the photon point source and the sensor. For each distance the results are obtained with the world object of PTFE (encapsulated scintillators) and the world of air (not encapsulated scintillators).

Figure 12 demonstrates that the majority of photons arrives during the first half nanosecond after generation. The onset of the photon signal shifts from zero at 0 source-to-sensor distance to 0.1 ns at 15 mm (half-height of the scintillator) and to 0.2 ns for the maximum possible distance of 30 mm. The rise time of the leading edge always remains fast (~0.01 ns).

Change of the world material from air to PTFE does not contribute to a strong increase in the number of collected photons (<15%). This is due to the fact that YAP has high refractive index, making a narrow YAP slab surrounded by a layer of air an effective lightguide due to full internal reflection. Since the benefit of adding encapsulation seems to be not significant and the results of the optical simulations strongly depend on the optical properties of the PTFE interface, in this study we have decided to consider only the model without encapsulation.

The obtained photon flight time values presented above have to be considered taking into account the emission properties of YAP: 0.38 ns rise time and 27/140 ns decay times [23]. Aiming at 1 ns waveform bins, the model can utilize the YAP emission time spectrum, delayed by a 0.1 ns per 15 mm of the source-to-window distance (see figure 13).

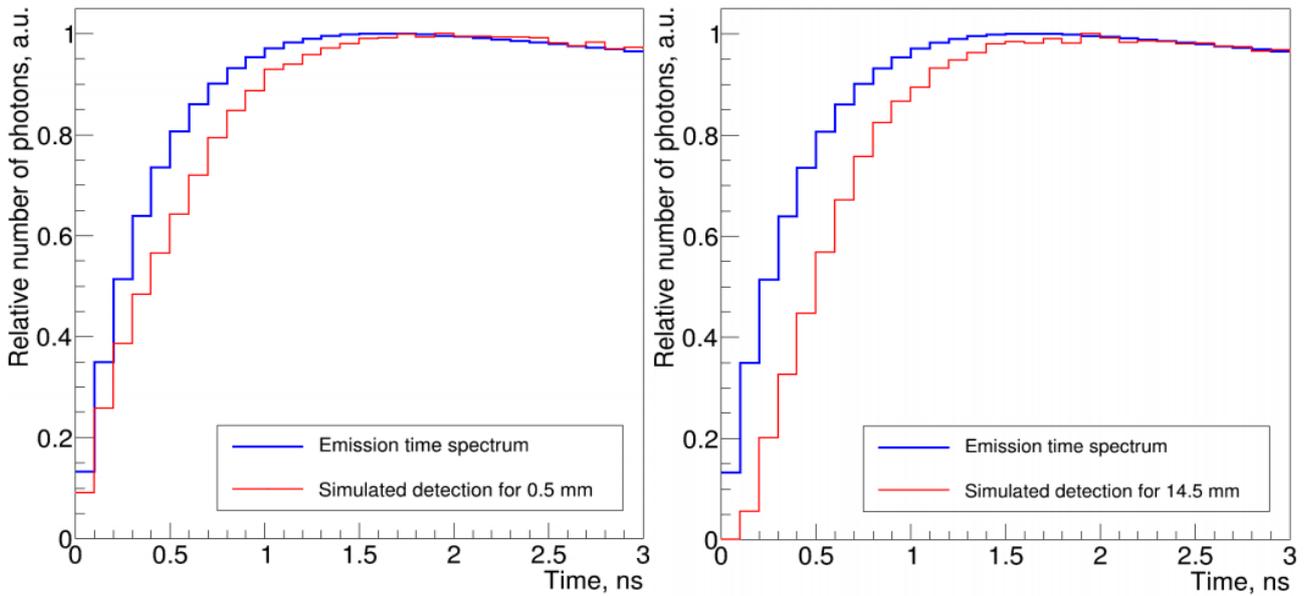

Figure 13. YAP emission (number of emitted photons per 0.1 ns) vs time from excitation in comparison with the number of detected photons (per 0.1 ns) vs time from emission, simulated for a 26 x 3 x 30 mm$^3$ YAP slab (see section 2.3). The data are shown for two source-to-sensor window distances of 0.5 mm (left) and 14.5 mm (right).

Our simulations show that the time spectra of the number of detected photons do not depend on the lateral coordinates of the point source inside the scintillator. Figure 14 shows that while the distance between the source position and the sensor window (Z position) remains the same, the detection curve vs time is essentially the same for all X and Y positions inside the scintillator. There is also a systematic decrease in the number of detected photons with distance between the photon source and the sensor window (see figure 15).

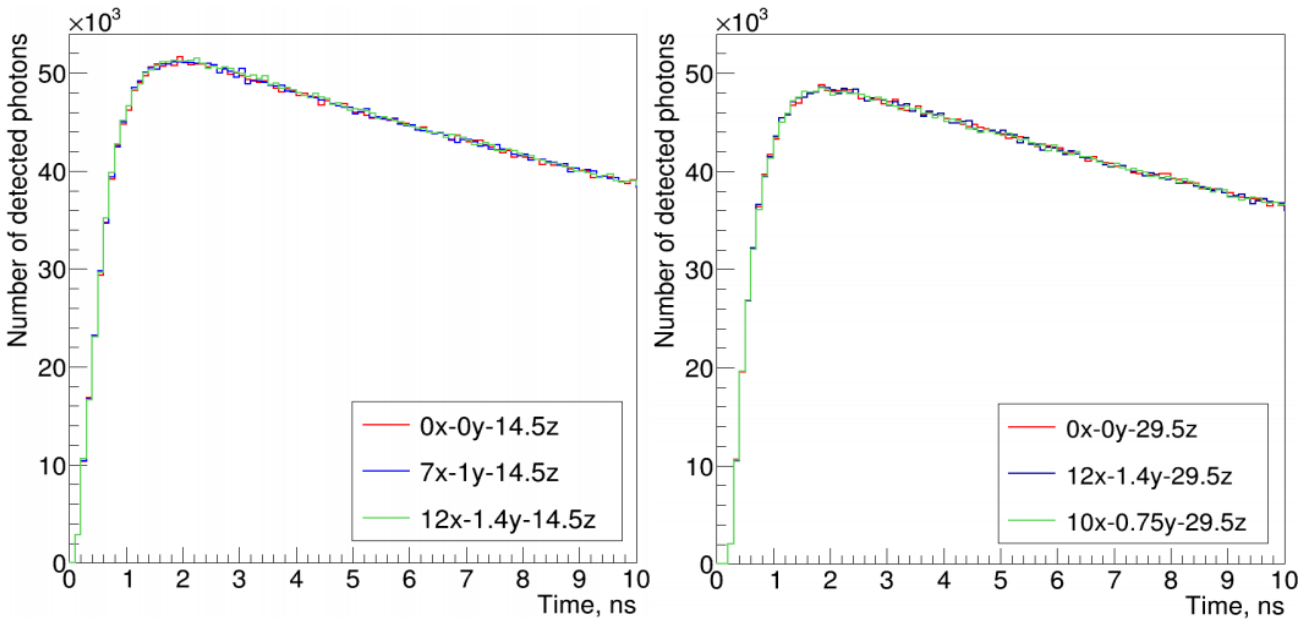

Figure 14. Number of detected photons (per 0.1 ns bin) as a function of X-Y-Z local coordinates of the source inside the slab in millimeters. X and Y are the source displacement from the scintillator's center line, and Z is the distance from the source to the sensor.

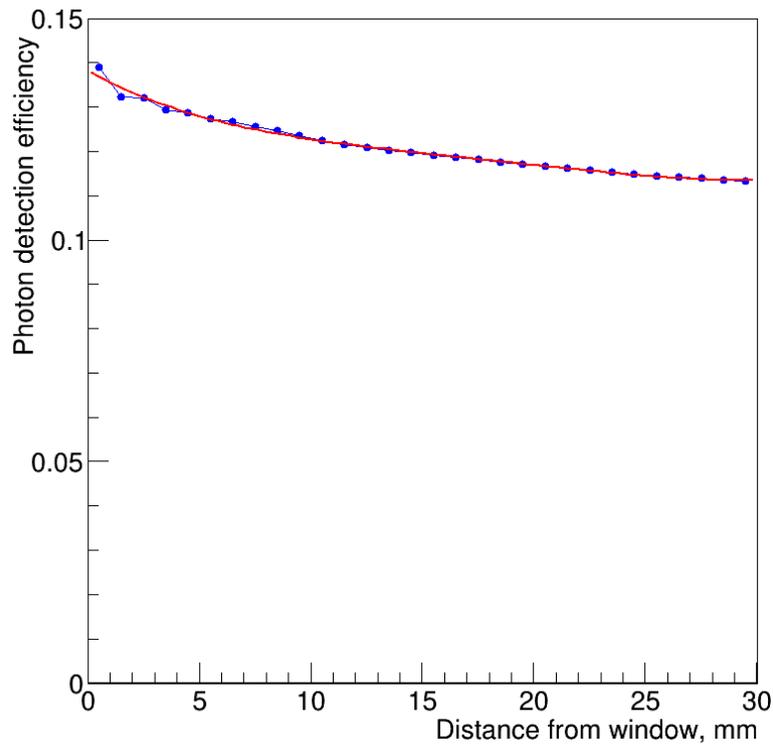

Figure 15. Photon detection efficiency vs distance between the source and the sensor's window (line with dots) and a 4th degree polynomial fit (red line).

Based on these results, the following model was developed which generates a bin-wise contribution to the signal waveform of a light sensor for a given deposition event, characterized by the deposited energy, the global time, and the distance between the deposition position and the sensor. First the YAP emission time spectrum with bins of 0.1 ns is scaled according to the deposited energy and the detection efficiency at the given source-to-sensor distance. Then the spectrum is shifted in time by one bin for the source-to-sensor distances in the range of 10 – 20 mm and by two bins for the distances of 20 – 30 mm. After that another time shift is applied to account for the delay between the global time of the deposition and the fixed time of start of the 1 ns bins of the signal waveform. Due to the difference in the bin spans (0.1 ns bins of the time spectrum and 1 ns bins of the signal waveform) this shift can have values from 0 to 9 bins. In the next step the bins of the time spectrum are accumulated in groups of 10 in order to calculate the expected values of the signal contributions to the bins of the signal waveform. Finally, the bin-wise values of the contributions are calculated using a random number generator with Poisson distribution and the corresponding expected values.

Figure 16 shows an example of comparison of the model output with the result of a full photon simulation for 14.5 mm distance between the deposition position and the sensor for several deposition-to-bin-start delays of 0, 0.3, 0.6 and 0.9 ns. The waveforms generated by the model are very similar to the simulated results. Good agreement was produced for all tested combinations of the distance and the time delay. The maximum difference in the integral of the generated waveform between the model and the simulation was less than 6 %.

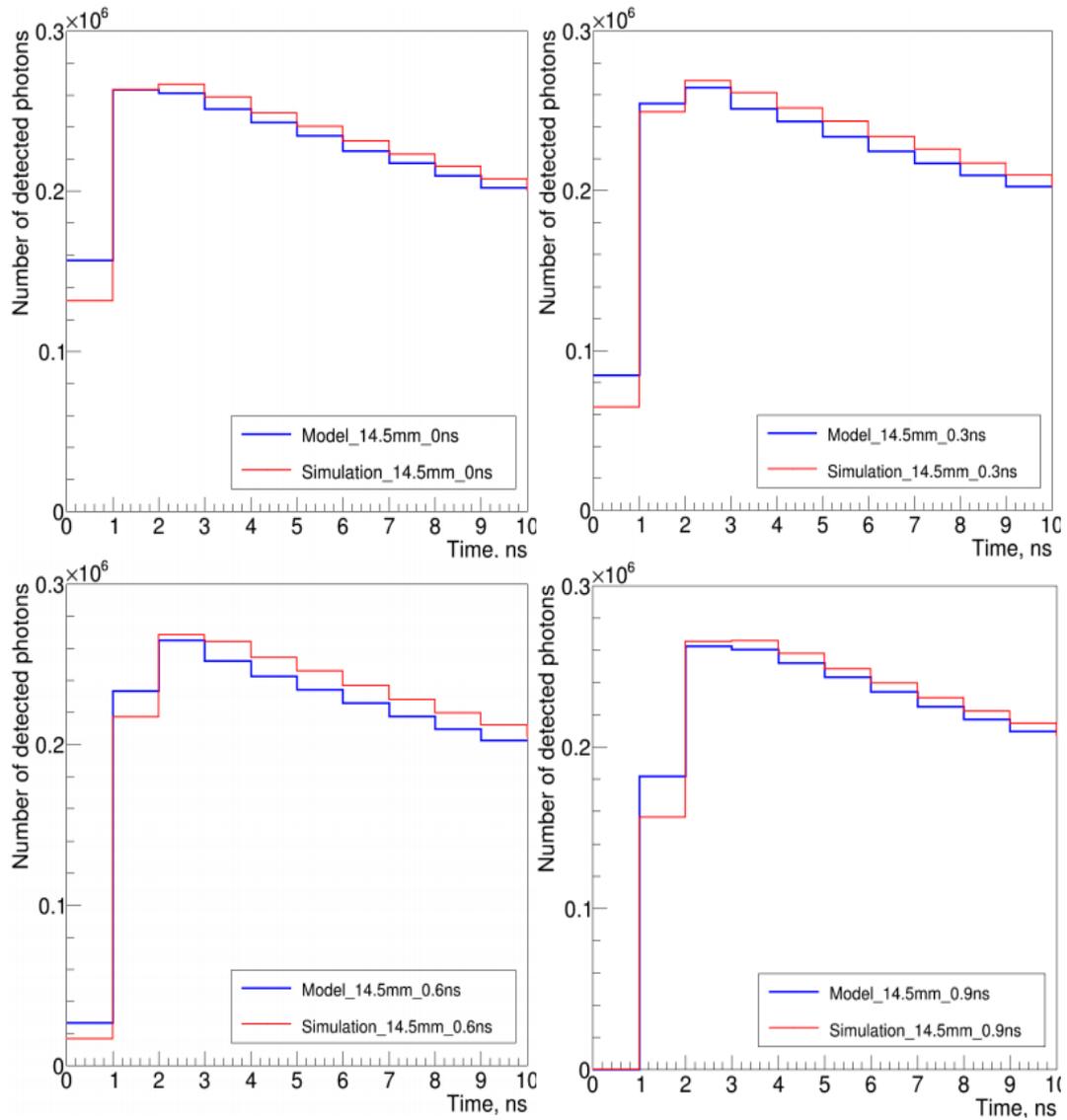

Figure 16. Number of detected photons vs time for the model output and for the results of an optical simulation. The data are given for 14.5 mm distance between the source and the sensor. The four graphs show the results for different time delays (the difference between the global time of the energy deposition and the beginning of the first bin of the waveform with the signal of the sensor) of 0, 0.3, 0.6 and 0.9 ns.

## 3.6 Waveform processing

Figure 17 shows an example of a signal waveform (the number of detected photons per 1 ns time bin as a function of the global time) generated for one of the light sensors. The waveform exhibits quite low level of pile-up. The energy deposition of 750 keV (the energy threshold introduced in section 3.2) corresponds to a peak with a maximum of about 50 detected photons per bin.

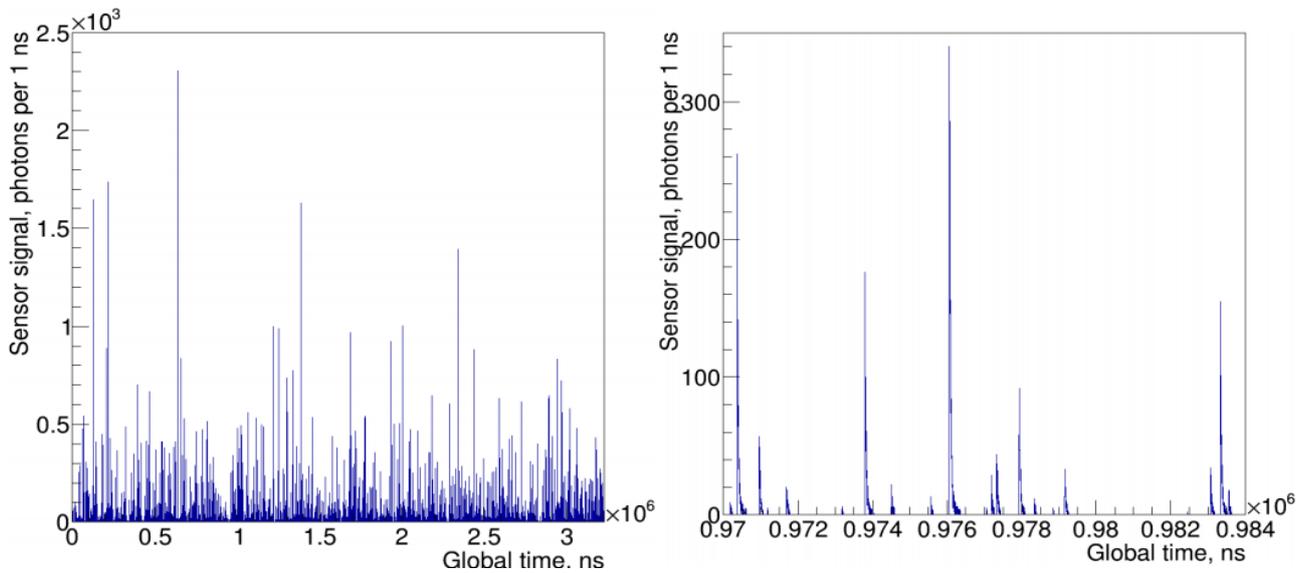

Figure 17. Example of a signal waveform: full time scale of 3.23 ms on the left and a zoom with the time range of 14 μs on the right.

Peak finding was performed taking advantage of the fast rise time of the signal from the optical sensors in response to a scintillation event: the rising part of the peak is always confined within 3 or less bins.

The peak finding algorithm has two states: 1) waiting for a new peak and 2) collecting peak data. A scan of the waveform starts at the first bin of the waveform. A peak candidate is marked if the signal in the current bin is at least 15 detected photons larger than the signal in the previous bin. The candidate is rejected if the difference between the current bin signal and the average signal over the last 3 bins is less than 3 standard deviations of the current signal value in order to avoid triggering on statistical fluctuations or on a weak peak superimposed over the tail of another peak.

If the candidate is accepted, the algorithm switches to the collecting state, in which the signal values are accumulated for the first 15 bins of the peak. According to our simulations, the cumulative signal during the first 15 ns corresponds to 40% of the total signal.

While in the collecting state, the algorithm still considers new peak candidates with every new bin. If another peak is detected in the peak collection state (possible pile-up event), the decision how to proceed is made in the following way: if the new peak is significantly stronger than the original one (at least 5 times larger in the sum signal of the first 3 bins of the peaks), the original peak is disregarded, and the peak collection restarts with the new one. In this situation the energy data extracted from the new peak are "contaminated" by the original peak over which the new peak is superimposed with, however the new peak is strong enough to consider this contamination negligible.

Otherwise, if the new peak is weak, the original and the new peaks are rejected since both time and/or energy information are not reliable for the two peaks and the algorithm is switched to the waiting state. Note that a new peak candidate in the collecting state is ignored if it appears within 3 bins from the start of the original peak to avoid re-triggering on the same rising slope.

An example of the test of the peak extraction algorithms is given in figure 18 and table 3. The results demonstrate good accuracy in the extracted energy and time bin index. The pile-up effect is properly taken into account: both peaks at 500 and 510 bin indexes are disregarded while from the peak pair at 800 and 810 only the much stronger peak at 810 is accepted.

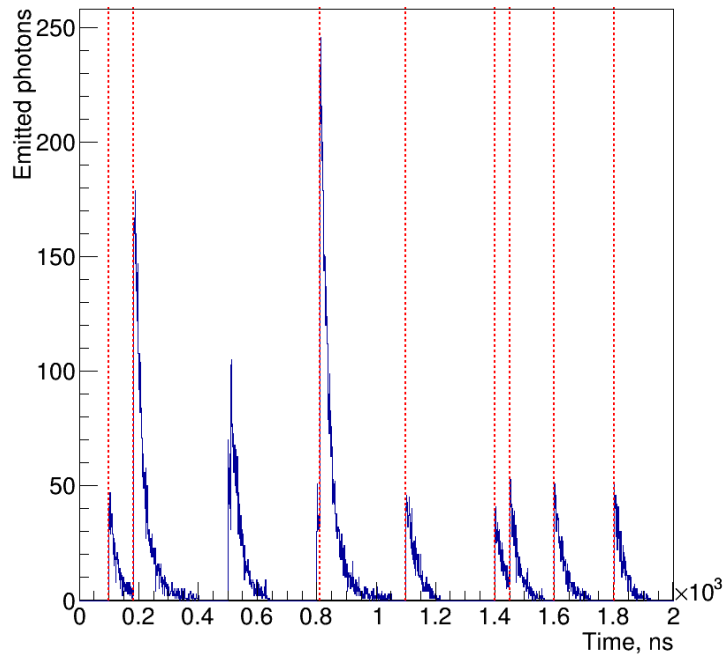

Figure 18. Waveform processing results. The red dotted lines indicate the positions of the accepted peaks.

| Values used to generate the waveform | | | Extracted | |
|---|---|---|---|---|
| Energy, keV | Time bin, # | Time offset, ns | Energy, keV | Time bin, # |
| 550 | 100 | 0 | 586 | 100 |
| 2500 | 183 | 0 | 2636 | 183 |
| 750 | 500 | 0 | - | - |
| 750 | 510 | 0 | - | - |
| 550 | 800 | 0 | - | - |
| 3000 | 810 | 0 | 3450 | 810 |
| 550 | 1100 | 0 | 669 | 1100 |
| 150 | 1110 | 0 | - | - |
| 550 | 1400 | 0 | 537 | 1401 |
| 550 | 1450 | 0 | 670 | 1450 |
| 650 | 1600 | 0.3 | 649 | 1600 |
| 650 | 1800 | 0.7 | 682 | 1801 |

Table 3. Example of the test results for the procedure used to extract the deposition data from a signal waveform.

The most conclusive test of the developed waveform processing method is to compare spatial profiles of the deposited energy vs scintillator index obtained by this method with the ones computed directly from the energy deposition data known from the simulations (see section 3.2). Figure 19 shows a comparison of the profiles obtained for the same deposition data using these two methods. The filtering on the deposition time (accepted range is from 1 to 3 ns) and deposition energy (accepted only depositions above 750 keV) are applied in both cases.

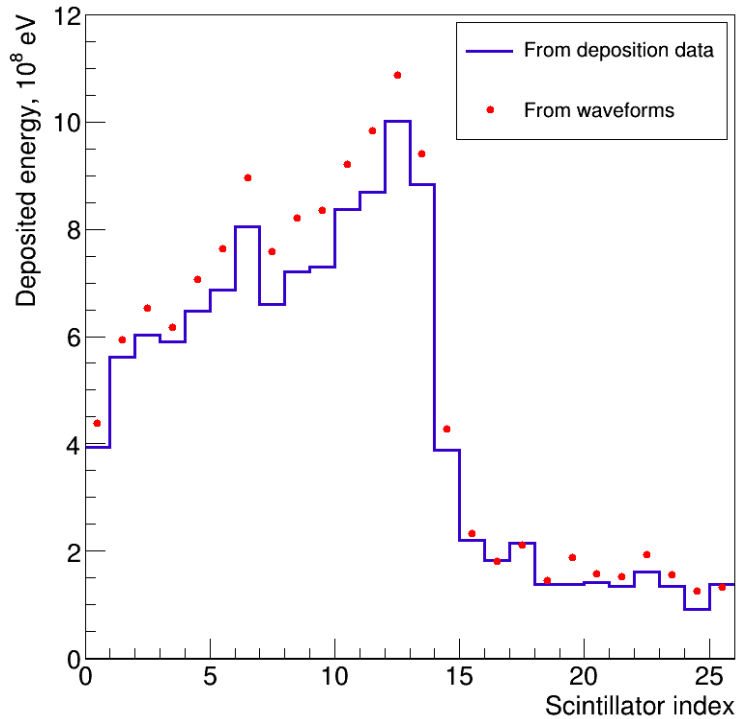

Figure 19. Comparison of the profiles of the deposited energy vs scintillator index obtained using deposition information available from simulations as well as utilizing the waveform generation/analysis procedure. The deposition data are filtered accepting only deposition events in the time window from 1 to 3 ns and with deposition energy above 750 keV.

The profiles show a very high degree of similarity. A slightly larger deposition energy provided by the waveform-based processing is most likely explained by the fact that signals of the light sensors depend on the distance between the deposition position and the sensor window. During the analysis of the waveform the information on this distance is not available, and the scaling of the extracted energy is performed assuming deposition in the middle of the scintillator slab (see figure 15). Another reason is due to the energy filtering. While the energy threshold is set to the same value, in the deposition analysis the filter is applied on the beam bunch basis (deposition in each scintillator is collected for one beam bunch and then the filter is applied), while in the waveform-based approach the filtering is more realistic, since depositions from several beam bunches, overlapping in time (e.g. a deposition from a prompt gamma ray from one bunch plus neutron-originating deposition from one of the previous bunches), are properly superimposed before applying the energy filter.

## 3.7  Precision of the edge determination

Based on the optimization results (sections 3.3 and 3.4), the following parameter values for the collimator were selected: the septal thickness of 3.2 mm, the height of 175 mm and the aperture of 3.8 mm for the configuration with one scintillator per aperture, and the septal thickness of 2.3 mm, the height of 260 mm and the aperture of 4.3 mm for the configuration with two scintillators per aperture.

Precision in the determination of the distal edge position was evaluated for both developed methods: using the deposition information (section 3.2) and based on the generation/processing of the sensor signal waveforms (section 3.6). The evaluation was performed in both cases using the same 120 files obtained in the phantom simulations with 12 different phantom shifts (see section 3.1). Each file was used in 4 independent runs of the second stage simulation (see section 3.2), providing in total 480 PT frames for evaluation of the edge determination accuracy.

The results obtained with both methods for the configuration with one scintillator per aperture are given in figure 20. The graphs on the left hand side show the extracted edge position as a function of the phantom shift, which is equal to the shift in the true position of the distal edge. The graphs on the right hand side show the distributions of the corrected edge position (the extracted edge position minus the phantom shift).

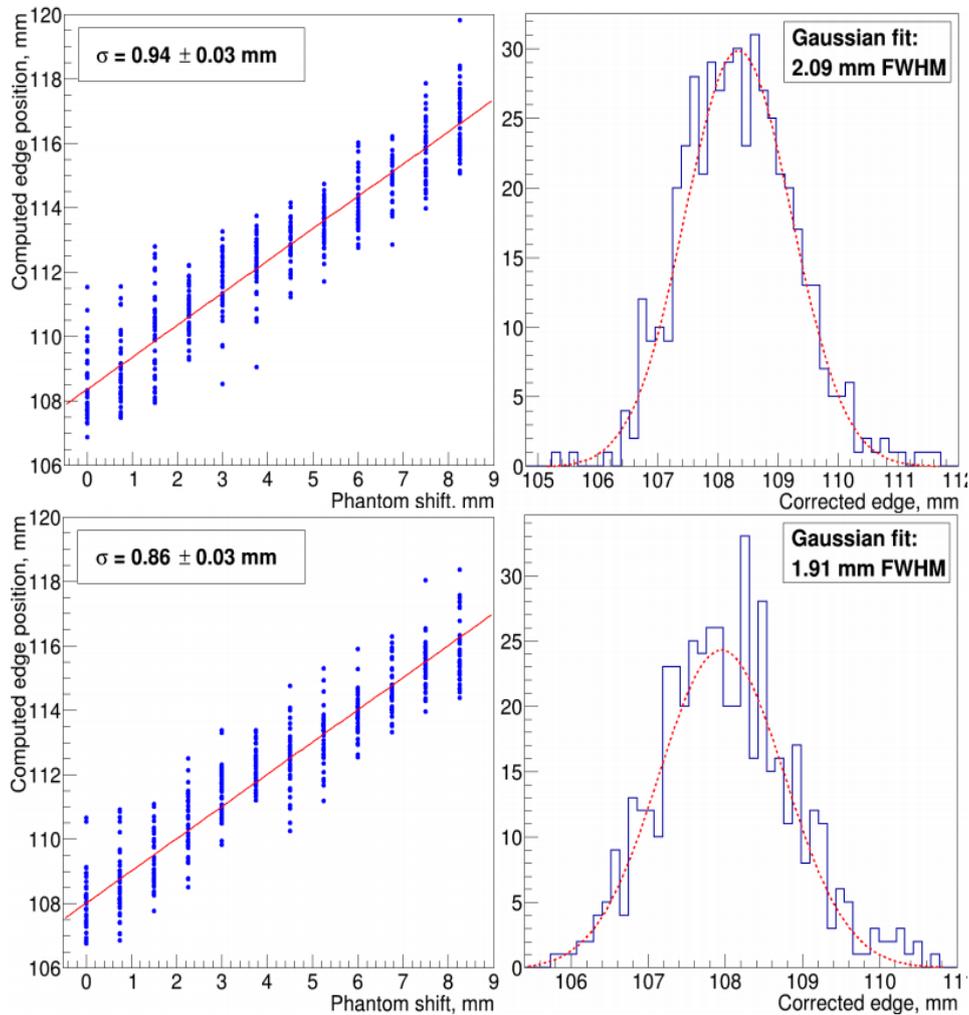

Figure 20. Edge position vs phantom shift (left column) and the distribution of the corrected edge position (right column) obtained using the deposition-based (top row) and the waveform-based (bottom row) data processing methods for the configuration with one scintillator per collimator aperture.

The standard deviation of the corrected edge position is equal to 0.94 ± 0.03 mm and 0.86 ± 0.03 mm for the deposition- and the waveform-based methods, respectively. The standard error in the standard deviations is defined by the number of samples (480) assuming normal distribution.

The results for the configuration with two scintillators per collimator aperture are shown in figure 21. The standard deviation of the corrected edge position is equal to 1.07 ± 0.03 mm and 1.10 ± 0.04 mm for the deposition- and the waveform-based methods, respectively.

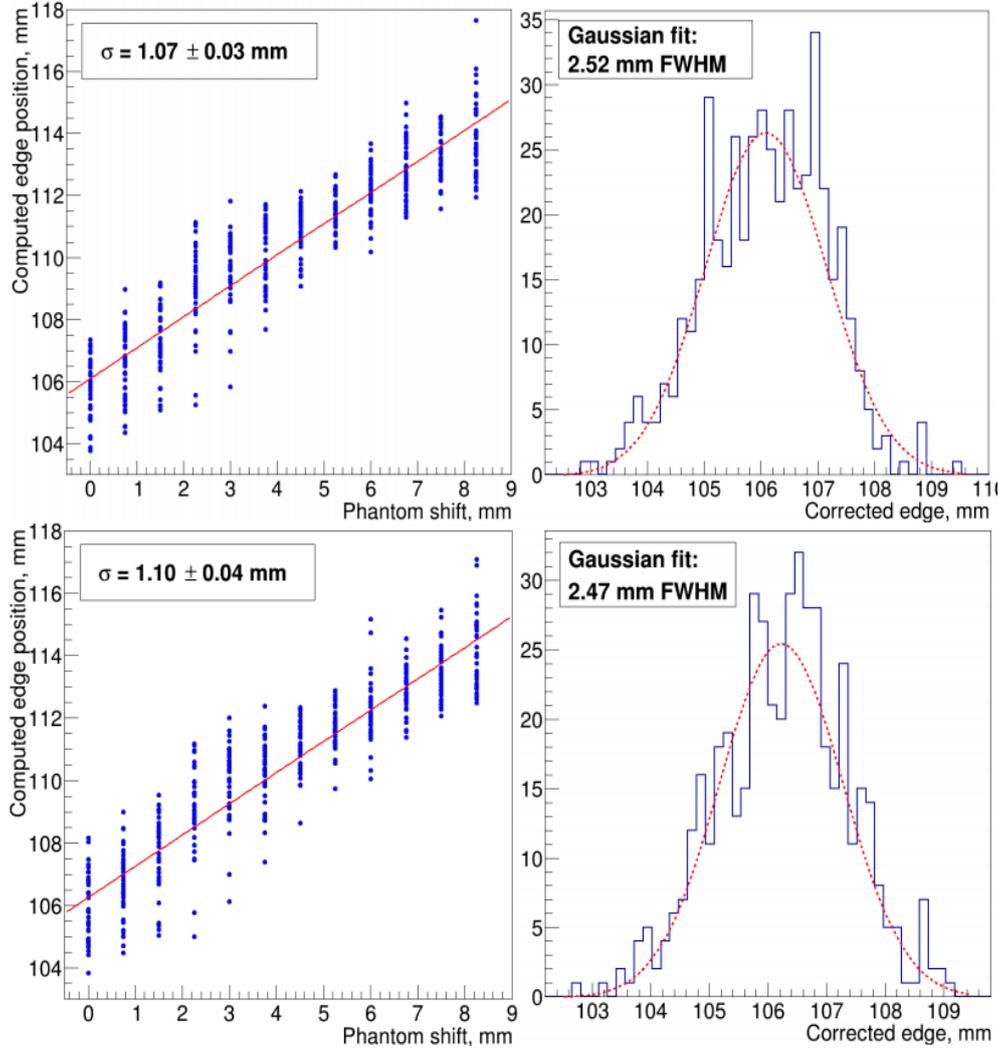

Figure 21. Edge position vs phantom shift (left column) and the distribution of the corrected edge position (right column) obtained using the deposition-based (top row) and the waveform-based (bottom row) data processing methods for the configuration with two scintillator per collimator aperture.

# 4  Discussion

The results of this study demonstrate that the multi-slat prompt-gamma camera allows to determine the position of the distal edge of the Bragg peak for 130 MeV proton beam in a PMMA phantom with a precision of about 2.1 mm FWHM. The edge determination precision does not depend on the relative position of the Bragg peak and the collimator elementary cell. The configuration with one scintillator per collimator aperture is performing ~20% better in comparison to the one with two scintillators. Considering the experimental prototype, since the latter configuration is more complex and requires more readout channels, the former one seems to be a better choice. One of the methods for the edge position determination applied in this study is based on analysis of the signal waveforms of the optical sensors, and thus can be used for analysis of experimental data without significant modifications.

The study had several assumptions which have to be addressed when considering design of an experimental prototype of the system. Electronic noise of the light sensors was not simulated. The noise can make processing of waveforms more challenging and lead to worsening of the precision. It was also assumed that the light sensors are directly coupled to the scintillators and they cover the entire lower surface of the scintilaltor plates. Considering practical limitations, usage of segmented sensors or implementation of lightguides might become necessary. Note that addition of a lightguide can have a significant broadening effect on the distribution of the photon flight time, and thus result in a reduction of efficiency of the time-based filtering of prompt gamma events.

Further optimization of the system is still possible. We have considered only one realistic distance between the phantom and the collimator of 150 mm and a fixed height of the scintillators of 30 mm. Different shapes of the collimator blades can also be implemented (e.g., trapezoidal). We have tested only two possible scintillator configuration, and in both of them the scintillators were installed between the collimator plates. We have preliminary results for a configuration with a monolithic scintillator situated behind the collimator, but due to spread of light over a large number of sensors and long photon travel times, the performance of this configuration is significantly worse in comparison to the configurations with scintillators installed in the collimator apertures.

In this study we have evaluated the performance of the camera without encapsulation of the scintillators. Our results show that addition of a PTFE layer does not significantly improve the photon collection efficiency since narrow scintillator plates with high refractive index exhibit strong light-guiding effect. However, the applied model of light scattering was quite primitive, and an experimental study is necessary to determine the benefits of utilizing different encapsulation schemes with Lambertian and specular reflectors.

More elaborate methods of the edge position determination from the deposition profiles compared to the one applied here can be considered, using, e.g., signals of a larger number of sensors. In this study, due to the optimization efforts, it was crucial to develop a robust automatic method, capable to provide meaningful data in a broad range of the camera's geometries.

We plan to continue the camera design optimization, in particular, to analyze the influence of the scintillator height, which was fixed in this study based on practical limitations. We also consider adding a neutron moderator/attenuator in front of the tungsten collimator in order to reduce the neutron-related background. A larger parameter space for these studies makes optimization significantly more computationally expensive, which can be addressed by implementing the semi-automatic optimization, applied in this study, on a grid of computers.

In the ongoing follow-up study we will extend the developed approach to 200 MeV proton beams (a typical energy for the pelvis area in PT) and replace the PMMA phantom with a realistic anthropomorphic one.

# 5 Conclusion

The results of this simulation study suggest that with a multi-slat prompt-gamma camera it is feasible to determine the position of the distal edge of the Bragg peak for a 130 MeV proton beam in a PMMA phantom with a precision of about 2.1 mm FWHM (σ < 1.0 mm). The simulations were performed considering particle transfer in the phantom and the camera as well as generation, transport and detection of optical photons.

One of the methods for the edge position determination, developed in this study, is based on analysis of the signal waveforms of the optical sensors, and, therefore, can be applied for analysis of the experimental data. The performance of the method was cross-compared with that of a method directly using energy deposition data available in simulations. The results obtained with the two methods are very similar.

Two schemes of scintillator packing were considered, the first with one and the second with two scintillators per collimator aperture. For both schemes, the collimator design optimization targeting the highest possible precision in the edge position determination was performed with the collimator septal thickness, height and aperture size as free parameters. The edge extraction precision is better (~20%) for the configuration with a single scintillator per collimator aperture.

# 6 Acknowledgments

This work was financed by the Portuguese national funds via FCT - Fundação para a Ciência e a Tecnologia, I.P., in the context of the project CERN/FIS-TEC/0019/2019. The authors acknowledge the Laboratory for Advanced Computing (*https://www.uc.pt/lca*) at the University of Coimbra for providing computing resources that have contributed to the research results reported in this paper.